\documentclass[pdflatex,sn-mathphys-num]{sn-jnl}

\usepackage{graphicx}%
\usepackage{multirow}%
\usepackage{amsmath,amssymb,amsfonts}%
\usepackage{amsthm}%
\usepackage{mathrsfs}%
\usepackage[title]{appendix}%
\usepackage{xcolor}%
\usepackage{textcomp}%
\usepackage{manyfoot}%
\usepackage{booktabs}%
\usepackage{algorithm}%
\usepackage{algorithmicx}%
\usepackage{algpseudocode}%
\usepackage{listings}%

\usepackage{tcolorbox}
\usepackage{multirow} 
\usepackage{url} 
\usepackage{subcaption}
\usepackage[dvipsnames]{xcolor}
\usepackage{siunitx}

\theoremstyle{thmstyleone}%
\theoremstyle{thmstyletwo}%

\theoremstyle{thmstylethree}%

\begin{document}

\title[Article Title]{One Model, Many Skills: Parameter-Efficient Fine-Tuning for Multitask Code Analysis}


\author*[1]{\fnm{Amal} \sur{Akli}}\email{amal.akli@uni.lu}

\author[2]{\fnm{Maxime} \sur{Author}}\email{maxime.cordy@uni.lu}

\author[3]{\fnm{Mike} \sur{Author}}\email{michail.papadakis@uni.lu}

\author[4]{\fnm{Yves} \sur{Le Traon}}\email{Yves.LeTraon@uni.lu}

\affil[1,2,3,4]{\orgname{University of Luxembourg}, \orgaddress{\country{Luxembourg}}}

\abstract{Large language models (LLMs) such as GPT-4 have recently surpassed specialized systems on code generation, yet their effectiveness on other code-analysis tasks remains less clear. At the same time, multi-task learning (MTL) offers a way to unify diverse objectives within a single model, but fully fine-tuning LLMs across tasks is computationally prohibitive. Parameter-efficient fine-tuning (PEFT) mitigates this cost by updating only a small fraction of weights. Although PEFT has proven effective in single-task settings, its potential for multi-task learning has not yet been systematically explored.

We present the first comprehensive evaluation of multi-task PEFT for code analysis, comparing several methods across diverse tasks and model architectures. Our experiments show that a single PEFT module shared across tasks can match, and in some cases surpass, full multi-task fine-tuning, confirming that the benefits of PEFT extend beyond isolated tasks. When comparing single-task and multi-task setups, we find that multi-task PEFT achieves a favorable performance-efficiency trade-off: it delivers accuracy close to single-task fine-tuning while reducing storage requirements, cutting the number of trainable parameters by a factor of the task count, and lowering computation costs by as much as 85\%. At the same time, multi-task gains remain sensitive to task grouping. Through task-pairing experiments, we identify key factors shaping outcomes: task stability, model architecture, task complementarity, asymmetry, and dataset quality determine the success of co-fine-tuning.

Finally, we benchmark efficient multi-task PEFT against direct prompting of open-source general-purpose LLMs, including DeepSeek, Qwen, Mistral, CodeLlama, and StarCoder. Despite their strong performance in code generation, these models underperform on classification and retrieval tasks, where even a 1B-parameter model with multi-task PEFT achieves significantly better results. This underscores the importance of parameter-efficient multi-task fine-tuning as a practical and effective alternative.

\vspace{2ex}
}


\maketitle

\section{Introduction}


Recent large language models have demonstrated significant advances in natural language processing and now perform well on a wide range of tasks, often cited as steps toward more Artificial General Intelligence (AGI). This progress extends to software engineering: on the \emph{LiveCodeBench} benchmark \cite{jain2024livecodebench}, for example, the continually updated leaderboard reports state-of-the-art generative models (e.g., O4-Mini) achieving \(\approx\!80.2\%\) \textsc{Pass@1} on code generation as of September~2025 \cite{lcb_leaderboard}, surpassing task-specific models. However, recent surveys and execution-based benchmarks suggest that \emph{code understanding} and \emph{program analysis} still lag behind pure generation \cite{hou2024llmse,cruxeval,codescope}.

To specialize LLMs for software engineering tasks \cite{lu2021codexglue}, one typically needs to fine-tune an LLM by updating its model parameters on smaller, task-specific datasets \cite{devlin2019bert}. Although traditional fine-tuning involves updating all model parameters, the size of recent models, often comprising billions of parameters \cite{codellama, openai2023gpt4, guo2024deepseek}, makes this process memory-intensive, computationally demanding, and challenging for both training and deployment in resource-constrained environments.


To address the challenges of full fine-tuning, parameter-efficient fine-tuning techniques have been proposed \cite{peft}. These methods involve training a small fraction of additional parameters, often less than 1\% of the total model parameters, while keeping the original model weights frozen. Prominent PEFT approaches include Adapters \cite{peft, PA}, Low-Rank Adaptation (LoRA) \cite{hu2022lora}, Prefix Tuning \cite{li2021prefix}, and variants of these methods such QLoRA \cite{dettmers2023qlora}. These techniques have demonstrated the ability to achieve performance comparable to full fine-tuning across various tasks, all while significantly reducing memory usage and computational requirements \cite{peft_survey,zou2023comprehensive,liu2023empirical,autoadapt}. Moreover, since only the additional parameters need to be stored, deploying multiple task-specific models becomes more storage-efficient, as the base model can be shared across tasks and loaded as needed. Previous work confirms that PEFT can match full fine-tuning on individual tasks, such as searching code of predicting defects \cite{autoadapt,peft_survey}. 

Similar to general-purpose models, Multi-Task Learning (MTL) combines multiple tasks into a single model, improving efficiency, reducing training costs, and memory needs \cite{mtl_survey}. Big companies like Facebook have adopted an intermediate pre-finetuning stage, essentially a large-scale multi-task learning step, between pretraining and task-specific fine-tuning to capitalize on shared representation learning \cite{aghajanyan-etal-2021-muppet}. 
Concurrently, Adapters have been shown to generalize across diverse programming languages without sacrificing performance \cite{adapter_all_langages}, and frameworks like MFTcoder \cite{mftcoder} have applied LoRA \cite{hu2022lora} and QLoRA \cite{dettmers2023qlora} to fine-tune together multiple tasks on code generation \cite{peft_codegen} with promising results. However, we still lack a systematic evaluation of how different PEFT methods behave when co-fine-tuning heterogeneous code tasks using multi-task learning. 


To address these gaps, we conduct the first systematic study to investigate the intersection of multi-task learning and parameter-efficient fine-tuning, aiming to specialize models that jointly learn several code-analysis tasks with greater efficiency and lower cost. Concretely, we benchmark multiple PEFT variants against full fine-tuning on a unified set of code-analysis tasks spanning diverse programming languages and model architectures. 
This analysis lets us answer the following research questions:

\textbf{RQ1}:  Are PEFT techniques effective in multi-task learning?



\textbf{RQ2}: What is the performance efficiency trade-off between multi-task and single-task PEFT?




\textbf{RQ3}: Which factors influence the performance of multi-task PEFT?


\textbf{RQ4} : How does multi-task PEFT compare with zero-shot prompting of general-purpose LLMs?

To answer those RQs, we fine-tune four code LLMs of different scales—UnixCoder \cite{guo2022unixcoder}, CodeT5+ Large \cite{wang2023codet5plus}, DeepSeek Coder \cite{guo2024deepseek}, and  Qwen2.5-Coder-1.5B \cite{qwen25coder},  across four distinct code-analysis tasks derived from the CodeXGLUE benchmark \footnote{https://github.com/microsoft/CodeXGLUE}. These tasks include code search using the AdvTest dataset \ footnote {https://paperswithcode.com/sota/code-search-on-codexglue-advtest} (a subset of the CodeSearchNet \cite{husain2019codesearchnet}); vulnerability detection using the Devign dataset \cite{zhou2019devign}; clone detection with the BigCloneBench dataset \cite{clonedataset}; and test-flakiness prediction using the FlakeFlagger dataset \cite{flakeflagger} and represent diverse software engineering objectives. We jointly fine-tune each model on all four tasks and compare full fine-tuning with four prominent PEFT strategies: serial adapters, parallel adapters, LoRA, and prefix tuning. We analyze the performance–efficiency trade-off against single-task fine-tuning on separate tasks, and we examine the factors that influence the performance of multi-task PEFT. Finally, we compare against directly prompting recent large instruct LLMs, including DeepSeek-Coder \cite{guo2024deepseek}, Qwen-Coder \cite{qwen25coder}, CodeLlama \cite{codellama}, Mistral \cite{codestral}, and StarCoder \cite{li2023starcoder}, in their largest released versions (up to 34B parameters). 

Our study shows that PEFT is broadly effective for multi-task learning, often matching and occasionally surpassing full fine-tuning while collapsing several tasks into a single model. Serial adapters emerge as the most reliable choice, while LoRA is especially beneficial for retrieval-style objectives such as code search. Compared to single-task PEFT, multi-task PEFT yields large efficiency gains by dividing the number of trainable parameters by approximately the number of tasks. Crucially, we find that transfer dynamics play a central role: task stability, model architecture, task complementarity, and asymmetry determine the success of co-fine-tuning, and task addition is not always beneficial. Finally, we show that multi-task PEFT on compact code-specialized backbones consistently outperforms zero-shot prompting of much larger general-purpose LLMs on code analysis tasks, preserving near-SFT accuracy while substantially reducing storage and compute requirements. Together, these findings provide the first systematic evaluation of PEFT methods for multi-task code analysis, offer practical guidelines for pairing tasks and architectures, and position multi-task PEFT of small-scale models as a better alternative to general-purpose LLMs.

\section{Background}
\subsection{Code LLMs}

Using LLMs for software engineering \cite{llms_for_se}, where models specifically designed to work with code---referred to as code LLMs---can interpret source code as sequences or graphs to perform tasks such as predicting bugs and vulnerabilities \cite{llm_for_vulDetetct}, generating code from natural language descriptions \cite{code_generation}, searching for code source based on a description \cite{code_search}, translating between programming languages \cite{code_translate}, and more.
These models are generally grouped into three main types \cite{llms_for_se}: encoder-only models, which focus on understanding the relationships between words and their context and include examples like RoBERTa, CodeBERT, GraphCodeBERT, ALBERT, and DeepSeek Coder \cite{liu2019roberta,codebert,guo2021graphcodebert,lan2020albert,guo2024deepseek}; encoder-decoder models, which first encode the input into an internal representation and then decode it to produce new text, as seen in models like PLBART, UniXcoder, CodeT5, and CodeT5+ \cite{ahmad2021plbart,guo2022unixcoder,wang2021codet5,wang2023codet5plus}; and decoder-only models, which generate text directly from previous inputs.
This category encompasses the GPT series (from GPT-1 through GPT-4), OpenAI's Codex, Meta's CodeLlama, Salesforce's CodeGen \cite{openai2023gpt4,chen2021codex,codellama,nijkamp2022codegen}, and other state-of-the-art decoder-only models, including Llama 3 (Meta), StarCoder2 (BigCode), DeepSeek-Coder, Qwen2.5-Coder (Alibaba), and Codestral (Mistral) \cite{codellama, li2023starcoder, guo2024deepseek, qwen25coder, codestral}.

\subsection{Parameter efficient fine-tuning}

\begin{figure*}[!htbp]
\vspace{-1.0em}
  \centering
  \includegraphics[width=1\textwidth]{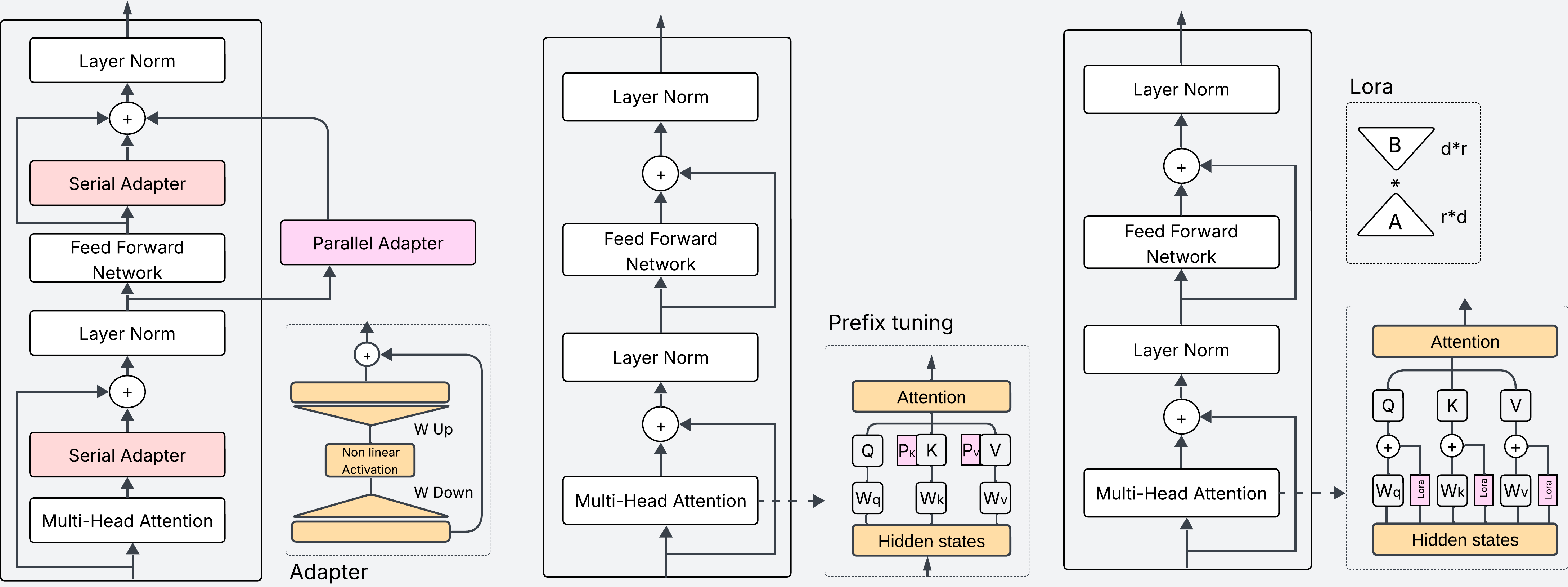}
  \vspace{-0.5em}
  \caption{Overview of four PEFT integration patterns in a Transformer block: serial adapters, parallel adapters, prefix-tuning, and LoRA. Colored components denote the added trainable modules, and dashed insets illustrate their internal layouts.}

  \label{fig:peft_methods}
\end{figure*}

Pre-trained models are first trained on large, multi-task datasets to acquire broad, general knowledge. They are then fine-tuned on a target dataset to specialize in a specific new task. Fine-tuning starts from the weights learned during pre-training and makes only minor updates, avoiding full retraining from scratch. Every parameter is updated in full fine-tuning, so the learning rate must remain low to prevent catastrophic forgetting.  With today’s LLMs this strategy is costly: modern LLMs contain billions of parameters such as GPT series \cite{openai2023gpt4}, CodeLlama \cite{codellama}, and DeepSeek \cite{guo2024deepseek} making full fine-tuning memory and compute intensive.
To address this, parameter-efficient fine-tuning methods have been proposed \cite{peft}. PEFT keeps all pre-trained weights frozen and learns only a small set of additional parameters such as Adapters \cite{peft}, Lora \cite{hu2022lora}, and prefix tuning \cite{li2021prefix}. PEFT techniques differ in where to insert these extra parameters and how many are required. 

The concept of adapters in Natural Language Processing (NLP) was first introduced by the \emph{Serial Adapter} framework \cite{peft}. As illustrated in Figure \ref{fig:peft_methods}, serial adapters are small Feed-Forward Network (FFN) blocks inserted within each Transformer layer. Each adapter consists of a down-projection matrix $W_{\mathrm{down}}$, followed by a non-linear activation function $\sigma(\cdot)$, and an up-projection matrix $W_{\mathrm{up}}$. Given a hidden input vector $h$, the adapter’s output is :

\begin{equation}
 Z = \text{W}_{\text{up}} (\sigma(\text{W}_{\text{down}}(h))) + h 
\end{equation}
\[  W_{\text{Up}} \in \mathbb{R}^{r \times d} \text{ and } W_{\text{down}} \in \mathbb{R}^{d \times r}  \]

$d$ denotes the hidden dimension of the transformer, and $r$ is the bottleneck dimension of the adapter.

In practice, each Transformer block is augmented with two adapter modules, one placed after the self-attention sublayer and the other after the FFN sublayer, following a sequential bottleneck design within the block. Although this design is straightforward, it can introduce additional serial dependencies that limit parallel execution. To address this, \cite{PA} proposed the \emph{parallel adapter} (Figure \ref{fig:peft_methods}), which restructures each adapter as a side network running in parallel with the original sublayer. 
Building on alternative PEFT strategies, \emph{Prefix-tuning} \cite{li2021prefix} introduces learnable vectors to the keys and values of attention in each layer (Figure \ref{fig:peft_methods}). More recently, \emph{LoRA} \cite{hu2022lora} rethinks weight updates by injecting low-rank matrices into the frozen projection weight matrix \(W\) as shown in Figure \ref{fig:peft_methods}.  \\


By introducing only a small number of trainable parameters, PEFT methods achieve performance comparable to full fine-tuning, while reducing the number of trainable weights by several orders of magnitude. Empirical studies show that training as little as \(\leq 1\%\) of the parameters can match or surpass the full fine-tuning, not only in natural language processing, but also in software engineering tasks such as code generation and analysis~\cite{liu2023empirical,zou2023comprehensive,peft_for_se_2025,autoadapt}. Moreover, PEFT has been found to better mitigate catastrophic forgetting~\cite{he-etal-2021-effectiveness}.

\subsection{Multi-task Learning}

Multi-task learning trains a single model on multiple related tasks simultaneously, leveraging both shared and task-specific information to boost learning efficiency and overall performance \cite{mtl_survey}. By sharing a subset of the model parameters across tasks, MTL introduces an inductive bias that improves generalization, mitigates overfitting, and often reduces computational cost compared to training separate models \cite{mtl_overview}. MTL approaches can be divided into two main types. \emph{Hard parameter sharing} \cite{Hardmultitask} fully shares the network backbone across all tasks, allowing a common representation to serve multiple objectives. In contrast, \emph{soft parameter sharing} \cite{soft_mtl} maintains separate models for each task while encouraging their parameters to be similar through regularization.

In the context of large language models, hard parameter sharing is especially attractive: the vast capacity of LLMs can accommodate multiple tasks within a single set of weights, enabling efficient multi-task adaptation without modifying the frozen backbone. Recent work has applied this principle to large‐scale pertaining. Aghajanyan et al. \cite{aghajanyan-etal-2021-muppet} introduce pre-finetuning (PFT) via the MUPPET framework, inserting an intermediary multi-task stage of approximately 50 tasks between pre-training and task-specific fine-tuning. Their experiments demonstrate consistent improvements in downstream performance and sample efficiency, provided at least 15 tasks are included. Building on this, Aribandi et al. \cite{aribandi2022ext5} propose ExT5, which pre-trains T5 on 107 supervised tasks; despite the risk of task interference, ExT5 outperforms the original T5 across a range of benchmarks.

In the context of parameter-efficient fine-tuning, Liu et al. \cite{mftcoder} introduced MFTCoder. This framework applies LoRA to jointly and in parallel fine-tune a single model across multiple code-generation tasks. Their experiments show that multi-task LoRA can match single-task fine-tuning (SFT) in certain settings. Building on this insight, we aim to evaluate whether PEFT methods, having demonstrated efficiency in single-task adaptation, can similarly support multiple tasks using only a small fraction of trainable parameters.
\section{Methodology}

Our primary objective is to assess how PEFT approaches can preserve task-specific performance when fine-tuning a shared model across multiple tasks. We adopt a joint training strategy in which a single model is trained simultaneously on all tasks. We further enhance the training process through dynamic loss weighting mechanisms that assign and optimize task-specific weights along with the model parameters.
We detail the data processing and preparation steps required to normalize heterogeneous datasets. Next, we describe the architectural choices in designing a multi-task model that incorporates shared components and task-specific heads. Finally, we outline the loss weighting strategies used to balance optimization across tasks with varying data sizes and complexities.

\subsection{Data processing}

\begin{figure*}[!htbp]
  \centering
  \includegraphics[width=\textwidth]{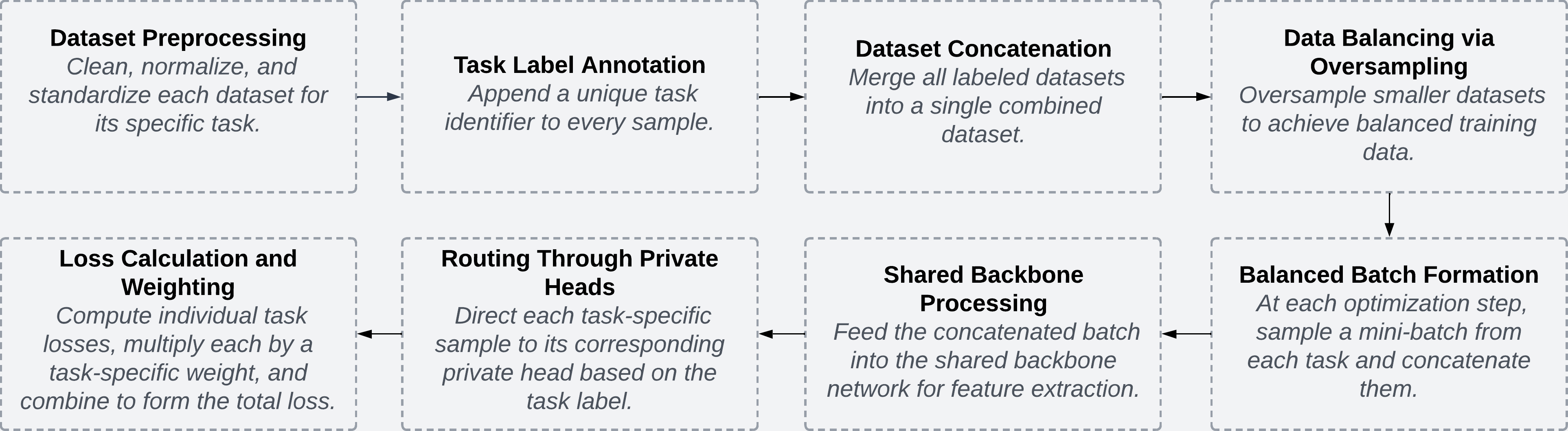}
  \caption{Data processing pipeline for our multi-task fine-tuning.}

  \label{fig:data_handling}
\end{figure*}

Figure \ref {fig:data_handling} summarizes the data processing steps. 
Initially, each task’s dataset is processed independently by tokenizing and filtering the inputs. All inputs are resized to a uniform length so that they can later be combined into a single batch. During data sampling, we assign a unique identifier to each task (e.g., task 0, task 1, etc.), ensuring that every sample comprises both the tokenized input and its corresponding task label. Subsequently, the datasets are concatenated using PyTorch’s “ConcatDataset” option.

The batch sampler is customized to iterate sequentially over the concatenated datasets to ensure that every training step receives a balanced contribution from each task. Specifically, it retrieves one batch per task in a round-robin way. In particular, when the sampler reaches the end of any individual dataset, it resets the index for that dataset, thereby restarting the sampling for that task, which effectively oversamples smaller datasets to maintain equal representation.  These batches are then merged into one global batch while maintaining a record of the order of the task-specific batches; this global batch is fed into the model. The shared encoder processes the entire global batch, after which, based on the task labels, each sub-batch is directed to its respective task head.

\subsection{Model design and train }

\begin{figure*}[!htbp]
\centering
\vspace{-1.0em}
\includegraphics[width=1\textwidth]{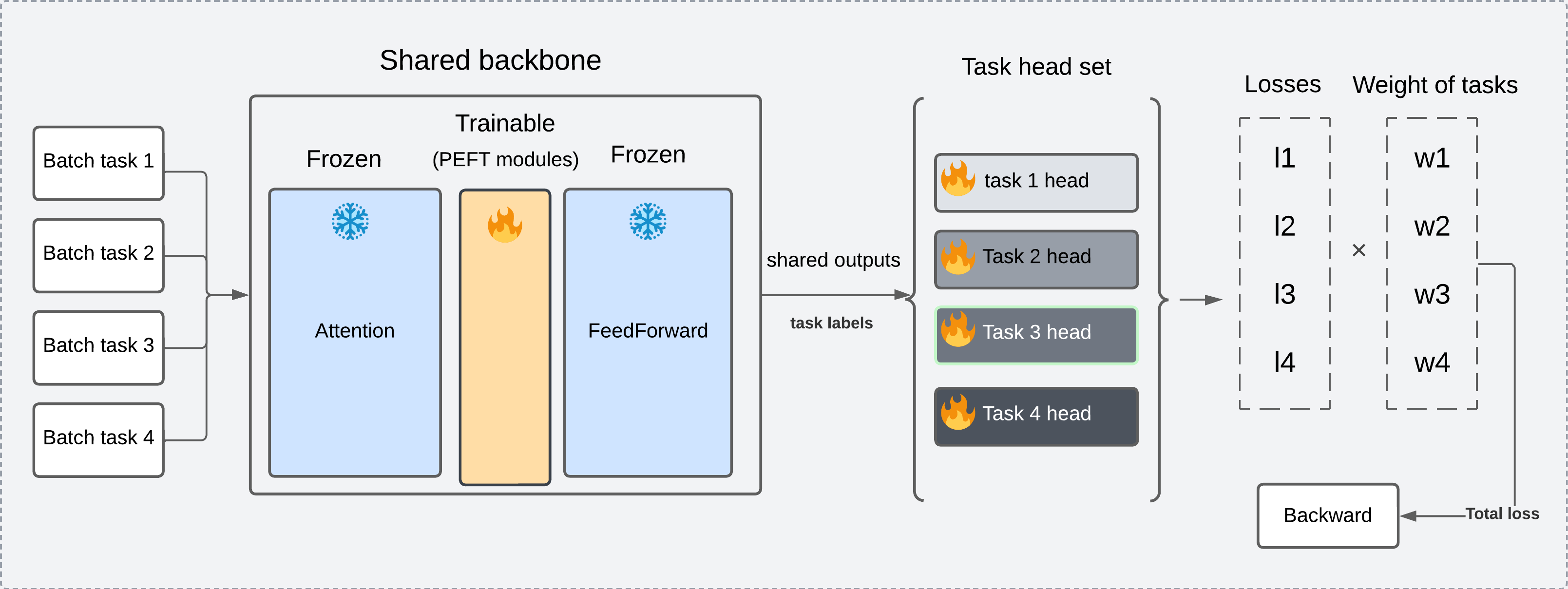}  
\vspace{-0.5em}
\caption{Overview of our PEFT-based multi-task fine-tuning pipeline.  At every optimization step, we draw a mini-batch from each task and pass the inputs through a \emph{shared backbone} whose original attention and feed-forward layers are \textcolor{blue}{\textbf{frozen}} (blue) while the inserted PEFT modules remain \textcolor{orange}{\textbf{trainable}} (orange).  The shared representation is routed to task-specific output heads, producing one loss \( \ell_{i} \) per task.  A set of learnable weights \( w_{i} \) balances these losses before they are summed and back-propagated through the PEFT blocks and the individual heads; backbone weights stay fixed.}

\label{fig:model_arch}
\end{figure*}

Hard parameter sharing is MTL's most widely employed strategy for deep learning and pre-trained models. In this approach, the encoder is shared among all tasks, while each task is assigned its dedicated head. Our experiments utilize a pre-trained encoder and append feedforward networks as task heads. These task heads are intentionally kept simple, typically comprising two projection layers with a non-linearity embedded within a dropout layer.

When employing parameter-efficient fine-tuning modules, such as Serial Adapter, Parallel Adapter, LoRA, and Prefix Tuning, we freeze the entire encoder and update only the inserted PEFT modules. In contrast, for full fine-tuning, all parameters are updated. We implement the PEFT modules using the open-source OpenDelta library \cite{hu2023opendelta}.
During training, we construct batches that contain an equal number of samples from each task. Each sample is accompanied by a task label, which redirects the encoder’s output to the appropriate task head, as shown in Figure~\ref{fig:model_arch}. Each task head produces its own output and computes an individual loss; these losses are then weighted, summed, and the combined loss is optimized with respect to the model parameters.
To adaptively balance multiple task-specific losses within our multi-task learning framework, we introduce a set of learnable weighting parameters, $\boldsymbol{\theta} = [\theta_1, \theta_2, \dots, \theta_K]$, where $K$ denotes the total number of tasks. Each parameter $\theta_k$ is initialized to zero (i.e., $\theta_k = 0$ for $k = 1,\dots,K$), ensuring that the initial weighting is uniform across tasks. These parameters are defined as \texttt{torch.nn.Parameter} objects in PyTorch and are updated alongside the model's other weights during backpropagation.
At each training iteration, we compute a normalized weight for each task by applying the softmax function:
\begin{equation}
\alpha_k = \frac{\exp(\theta_k)}{\sum_{i=1}^{K} \exp(\theta_i)},
\label{eq:softmax}
\end{equation}
where $\alpha_k$ represents the probability that the loss associated with task $k$ will contribute to the overall loss.

The overall loss function is then formulated as a weighted sum of the individual task losses:
\begin{equation}
\mathcal{L} = \sum_{k=1}^{K} \alpha_k \, \mathcal{L}_k,
\label{eq:total_loss}
\end{equation}
with $\mathcal{L}_k$ being the loss for task $k$. In this manner, the adaptive weighting mechanism is integrated directly into the training process. During backpropagation, gradients of the global loss $\mathcal{L}$ are propagated to both the base model parameters and the weighting parameters $\boldsymbol{\theta}$. Consequently, the model dynamically modulates the influence of each task based on the evolving training dynamics, thereby promoting effective learning and mitigating negative transfer.

\section{Experimental setup}
\subsection{Research questions}
Our study aims to answer the following research questions:

\subsubsection*{\textbf{RQ1. Are PEFT methods effective for multi-task learning?}}
We analyze whether PEFT can match full fine-tuning when a \emph{single} shared model is trained jointly on multiple tasks. 
We evaluate four PEFT strategies (LoRA, serial adapters, parallel adapters, and prefix tuning) on clone detection, code search, vulnerability detection, and test flakiness, using four public backbones that span scales (UniXcoder, CodeT5+ 770M, DeepSeek-Coder 1.3B, Qwen2.5-Coder-1.5B). 
We compare the performance per task of PEFT with full fine-tuning under the same joint training protocol.

\subsubsection*{\textbf{RQ2. What is the performance–efficiency trade-off between multi-task and single-task PEFT?}}
Under the same setup, we also fine-tune \emph{single-task} adapters and compare them to a \emph{multi-task} adapter trained once for all tasks. 
We analyze effectiveness and efficiency, quantifying the trade-off between single-task and multi-task PEFT.

\subsubsection*{\textbf{RQ3.  Which factors influence the performance of multi-task PEFT?}}
We evaluate all pairwise combinations of the four tasks across the four trained models using the Adapter and Lora methods, and our objective is to identify the important factors that influence the performance of multitask training and task combination (dataset size, task type similarity, programming language, model architecture and size, and PEFT method type). The outcome will translate into practical guidelines for designing effective multi-task pipelines for code analysis tasks.

\subsubsection*{\textbf{RQ4. How does multi-task PEFT compare with zero-shot prompting of general-purpose LLMs?}} We aim to evaluate the effectiveness of the latest large models on code analysis tasks compared to our multi-task PEFT method. 
We prompt large instruct models directly (CodeLlama, Qwen, DeepSeek-Coder, StarCoder, Mistral) with our unified prompts. We use the same evaluation protocol as in our fine-tuning, and compute the metrics on the same test splits and compare them to our multi-task PEFT and full fine-tuning results.

\subsection{Pre-trained models and LLMs}
To answer RQ1, RQ2, and RQ3, we need to train the models. 
To ensure that our results are not tied to a single backbone while staying within our hardware limits, we select four recent open-source code LLMs that cover the small and billion-parameter regimes, with different architectures (encoder–decoder and decoder-only).
For the small-tier encoder–decoder models we use \textit{UniXcoder-base}, pre-trained on 45M code–comment pairs (\(\sim\!125\)M parameters), and \textit{CodeT5+\,-Large} (\(\sim\!770\)M parameters), which extends CodeT5 with additional pre-training objectives and a larger multilingual corpus. 
To represent the billion-parameter class, we include the decoder-only models \textit{DeepSeek-Coder-1.3B} and \textit{Qwen2.5-Coder-1.5B}.

Given our GPU memory budget (46\,GB), these are the largest models we can train while keeping \emph{full fine-tuning} as one of our baselines; for example, a 7B model typically requires on the order of \(\sim\!80\)\,GB for full fine-tuning.

To answer RQ4 (direct inference of large instruction-tuned LLMs), we evaluate recent large instruction-tuned models using our unified prompts: \textit{DeepSeek-Coder-33B-Instruct}, \textit{CodeLlama-34B-Instruct-hf}, \textit{Qwen2.5-Coder-32B-Instruct}, \textit{StarCoder2-15B}, and \textit{Mistral-7B-Instruct-v0.3}.
We do not include paid APIs (e.g., GPT-4) because we evaluate large datasets (e.g., clone detection and code search with over a million samples), making API-based evaluation prohibitively expensive.

\subsection{Tasks and Datasets}
Our study targets \emph{code analysis} tasks only, since for code generation, recent large LLMs tend to outperform fine-tuned models. We include typical CodeXGLUE tasks covering binary classification and code retrieval. These tasks share a compatible model I/O format, which makes it possible to share most parameters across tasks, the main objective of hard parameter sharing in MTL.

\paragraph{Defect detection.}
Given a single C function, the classifier must decide whether the code contains a security vulnerability.  
We use the \textbf{Devign} corpus~\cite{zhou2019devign}, released with CodeXGLUE\footnote{\url{https://github.com/microsoft/CodeXGLUE/tree/main/Code-Code/Defect-detection}}, consisting of \(\approx 27{,}400\) functions from \texttt{FFmpeg} and \texttt{QEMU}. Each sample is labeled \emph{vulnerable} or \emph{safe} based on expert analysis of CVE reports. We follow the CodeXGLUE split: \(21{,}854\) train, \(2{,}732\) validation, and \(2{,}732\) test.

\paragraph{Clone detection.}
The task is to determine whether two Java methods implement the same semantics.  
We use \textbf{BigCloneBench}~\cite{clonedataset} via CodeXGLUE\footnote{\url{https://github.com/microsoft/CodeXGLUE/tree/main/Code-Code/Clone-detection-BigCloneBench/dataset}}.  
Because the original benchmark exceeds \(1.7\)M method pairs, we adopt the processed split: \(90{,}102\) training pairs and \(41{,}541\) pairs each for validation and test.

\paragraph{Code search.}
Given a natural-language query, the model must retrieve the most relevant code snippet from a large corpus.  
We evaluate on the \textbf{AdvTest} split of CodeSearchNet provided by CodeXGLUE\footnote{\url{https://github.com/microsoft/CodeXGLUE/tree/main/Text-Code/NL-code-search-Adv}}.  
AdvTest anonymizes identifier names in the test set, forcing models to rely on semantics rather than lexical overlap. The split contains \(251{,}820\) training queries, \(9{,}604\) validation, and \(19{,}210\) test.

\paragraph{Flakiness prediction.}
The goal is to predict whether a unit test will behave non-deterministically (flaky) or consistently.  
Experiments use the \textbf{FlakeFlagger} dataset~\cite{flakeflagger}, which includes \(21{,}802\) Java tests from open-source projects, labeled via repeated executions on CI infrastructure: \(13{,}081\) train, \(4{,}360\) validation, and \(4{,}361\) test.




   

 

\subsection{Metrics}

For all subsequent comparisons, we adopt the standard metrics widely used in literature and public leaderboards. 
Specifically, we use the F1 score for clone detection and flakiness prediction, as these datasets are imbalanced and F1 better reflects performance on both positive and negative classes \cite{clonedataset,flakeflagger}. 
For vulnerability detection, we report accuracy, which is the conventional metric in the Devign benchmark \cite{zhou2019devign}. 
For code search, we follow CodeXGLUE \cite{lu2021codexglue} and related works \cite{husain2019codesearchnet,jain2024livecodebench}, reporting mean reciprocal rank (MRR), which captures ranking quality and is the de facto standard for retrieval tasks.

\subsection{Prompt design}

To address RQ4, we designed basic prompts for instruction-tuned LLMs in accordance with OpenAI’s prompt engineering best practices. Specifically, we assign an explicit role to the system, pose a single, unambiguous question, and constrain the output format. For classification tasks, the prompt requires a binary decision (“Yes” or “No”), consistent with previous work \cite{prompt_vul, prompt_clone}. For code search, we use a prompt-based re-ranking scheme: given the query and the top-K retrieved snippets, the LLM returns a numeric relevance score (0–100) for each snippet with no explanation. We classify candidates according to these scores and compute MRR in the final ranking, making the prompt method directly comparable to the fine-tuned baseline.

\begin{tcolorbox}[title={Prompts used for different tasks }, colback=gray!5, colframe=gray!60]
\ttfamily\small
I want you to act as a \textbf{vulnerability detection} system. 
Is this code vulnerable? 
\textbf{CODE:}  \textit{<insert code snippet here>}  

Respond with \textbf{YES} or \textbf{NO}. 

\medskip\hrule\medskip

I want you to act as a \textbf{test flakiness} detection system.

Is this test code \textbf{ FLAKY}?

\textbf{TEST CODE:}  \textit{<insert test code here>} 

Respond with \textbf{YES} or \textbf{NO}. 

\medskip\hrule\medskip

I want you to act as a \textbf{code clone detection} system. 

Are the two snippets \textbf{SEMANTIC CLONES}?

\textbf{SNIPPET A:}  \textit{<insert snippet A here>} 
\quad

\textbf{SNIPPET B:}  \textit{<insert snippet B here>} 

Respond with \textbf{YES} or \textbf{NO}. 

\medskip\hrule\medskip

I want you to act as a \textbf{code search re-ranking} system. 

Given the natural-language query and candidate code snippets, assign a relevance score in \textbf{[0,100]} for \textbf{each} candidate (100 = exact implementation, 0 = unrelated). \\
\textbf{QUERY:} \textit{<insert natural-language query here>} -\\
\textbf{CANDIDATES:}\\
\textbf{ID}=1\ \ \textbf{CODE:}  \textit{<code snippet 1>} \\
\textbf{ID}=2\ \ \textbf{CODE:}  \textit{<code snippet 2>} \\
\textit{...}\\[2pt]
Output One line per candidate: \texttt{<ID>\textbackslash t<score>}. 
\end{tcolorbox}

\subsection{Implementation details}
All data and scripts used in this work are provided. \footnote{\url{https://github.com/Amal-AK/multitask_PEFT}} 
We apply the same hyperparameter schedule to every backbone to ensure comparability across tuning methods.  PEFT variants are trained with a learning rate (LR) of \(1\times10^{-4}\), whereas full fine-tuning uses \(2\times10^{-5}\), values commonly adopted in prior work. Optimization is performed with Adam (\(\beta_{1}=0.9,\ \beta_{2}=0.999\)), a maximum input sequence length of 512 tokens, and a global seed of 42.  The batch size is 32 for UniXcoder and 8 for CodeT5p-Large, DeepSeek-Coder, and Qwen-coder. Training proceeds for up to 10 epochs with early stopping on validation loss. 

\paragraph{Multi-task heads.}
Each classification task employs a two-layer MLP (hidden dimension \(d_{\text{model}}/2\) followed by a size-1 output) with ReLU activation.  For code search, we use a \(d_{\text{model}}\!\rightarrow\!512\) linear projection to produce fixed-width embeddings used in an in-batch retrieval loss.

\paragraph{PEFT configurations.}
We instantiate serial adapters, parallel adapters, LoRA, and prefix tuning through the \textsc{OpenDelta} library\footnote{\url{https://github.com/thunlp/OpenDelta}}.  Adapters use a bottleneck of 64 and are inserted in both attention and feed-forward blocks; LoRA applies rank-16 updates to the attention projections; prefix tuning retains library defaults (prefix length 20).  Some recent architectures (e.g.\ DeepSeek-Coder, Qwen-coder) are only partially supported by the current \textsc{OpenDelta} release, so not every PEFT variant could be evaluated on that backbone.

\paragraph{Hardware.}
All experiments were run on a single Linux server equipped with an Intel Xeon Silver 4416+ CPU and four NVIDIA L40S GPUs (46 GB each).
\section{ Experimental Results}

\subsection{RQ1: Multi-Task PEFT Fine-Tuning versus Multi-Task FULL}

\newcommand{\UP}[1]{\textcolor{green!60!black}{\scriptsize$\uparrow$\,#1}}
\newcommand{\DOWN}[1]{\textcolor{red!70!black}{\scriptsize$\downarrow$\,#1}}
\newcommand{\EQ}[1]{\textcolor{gray!60!black}{\scriptsize$\leftrightarrow$\,#1}}

\begin{table*}[!htbp]
\vspace{-0.5em}
\caption{Side-by-side \textbf{MFT} (top) and \textbf{SFT} (bottom). Methods in rows; metrics in columns.
Scores are aligned; deltas (absolute p.p. vs \emph{Full} of the same block) are shown in separate columns.
Bold marks the better score between MFT and SFT for the same model+method+metric.}
\centering
\begingroup
  \setlength{\tabcolsep}{6pt}     
  \renewcommand{\arraystretch}{1.15}
  \fontsize{10pt}{12pt}\selectfont  

\resizebox{\textwidth}{!}{%
\begin{tabular}{
|l|l|
S[table-format=3.2]|
S[table-format=2.2]r|
S[table-format=2.2]r|
S[table-format=2.2]r|
S[table-format=2.2]r|
}
\hline
\textbf{Model} & \textbf{Method} & \textbf{Trainable\%} &
\multicolumn{2}{c|}{\textbf{Clone F1 (\%)}} &
\multicolumn{2}{c|}{\textbf{Vuln Acc (\%)}} &
\multicolumn{2}{c|}{\textbf{Flakiness F1 (\%)}} &
\multicolumn{2}{c|}{\textbf{Search MRR (\%)}} \\

\hline

\multicolumn{11}{|c|}{\bfseries MFT (multi-task)} \\ \hline

\multirow{5}{*}{\bfseries Unixcoder-base}
& Full     & 100.00 & 93.70 & & 61.35 & & 71.20 & & 32.94 & \\
& Serial adapter   &   1.85 & \textbf{93.79} & \UP{0.09} & \textbf{61.35} & \EQ{0.00} & \textbf{71.90} & \UP{0.70} & 33.49 & \UP{0.55} \\
& Parallel adapter &   0.93 & 92.29 & \DOWN{1.41} & 61.05 & \DOWN{0.30} &  68.65 & \DOWN{2.55} & 32.50 & \DOWN{0.44} \\
& LoRA     &   0.46 & 92.33 & \DOWN{1.37} & 60.43 & \DOWN{0.92} & 66.20 & \DOWN{5.00} & \textbf{34.81} & \UP{1.87} \\
& Prefix   &   6.43 & 90.85 & \DOWN{2.85} & 60.98 & \DOWN{0.37} & 65.38 & \DOWN{5.82} & 30.86 & \DOWN{2.08} \\
\hline

\multirow{5}{*}{\bfseries Codet5+ 770M}
& Full     & 100.00 & \textbf{93.68} & & 61.79 & & \textbf{72.24} & & 21.64 & \\
& Serial adapter   &   1.70 & 92.28 & \DOWN{1.40} & \textbf{63.47 }& \UP{1.68} & 69.80 & \DOWN{2.44} & 17.40 & \DOWN{4.24} \\
& Parallel adapter  &   0.85 & 92.24 & \DOWN{1.44} & 60.87 & \DOWN{0.92} & 70.00 & \DOWN{2.24} & 20.01 & \DOWN{1.63} \\
& LoRA     &   0.42 & 92.56 & \DOWN{1.12} & 62.12 & \UP{0.33} & 69.80 & \DOWN{2.44} & \textbf{22.21} & \UP{0.57} \\
& Prefix   &   6.43 & 89.23 & \DOWN{4.45} & 61.09 & \DOWN{0.70} & 68.65 & \DOWN{3.59} & 15.83 & \DOWN{5.81} \\
\hline

\multirow{5}{*}{\bfseries Deepseek coder 1.3B}
& Full     & 100.00 & 93.31 & & \textbf{60.53} & & 68.71 & & 30.70 & \\
& Serial adapter    &   0.98 & 92.57 & \DOWN{0.74} & 60.43 & \DOWN{0.10} & 69.69 & \UP{0.98} & 30.83 & \UP{0.13} \\
& Parallel adapter &   0.84 & \textbf{93.72} & \UP{0.41} & 60.40 & \DOWN{0.13} & \textbf{70.18} & \UP{1.47} & 33.20 & \UP{2.50} \\
& LoRA     &   0.24 & 92.31 & \DOWN{1.00} & 60.25 & \DOWN{0.28} & 67.13 & \DOWN{1.58} & \textbf{33.90} & \UP{3.20} \\
& Prefix   &    \multicolumn{1}{c|}{--} & \multicolumn{1}{c}{--} & & \multicolumn{1}{c}{--} & & \multicolumn{1}{c}{--} & & \multicolumn{1}{c}{--} & \\
\hline

\multirow{5}{*}{\bfseries Qwen2.5-Coder-1.5B}
& Full     & 100.00 & \textbf{92.86} & & \multicolumn{1}{c}{60.76} & & 68.31 & & 31.06 & \\
& Serial adapter   &   0.71 & 92.68 & \DOWN{0.18} & \multicolumn{1}{c}{\textbf{60.87} } & \UP{0.11} & \textbf{72.54} & \UP{4.23} & 29.80 & \DOWN{1.26} \\
& Parallel adapter  &   0.35 & 92.17 & \DOWN{0.69} & \multicolumn{1}{c}{58.64 } & \DOWN{2.12} &  72.08 & \UP{3.77} & 32.06 & \UP{1.00} \\
& LoRA     &   0.14 & 92.47 & \DOWN{0.39} & \multicolumn{1}{c}{57.39} & \DOWN{3.37} &  70.97 & \UP{2.66} & \textbf{31.43} & \UP{0.37} \\
& Prefix   & \multicolumn{1}{c|}{--} & \multicolumn{1}{c}{--} & & \multicolumn{1}{c}{--} & & \multicolumn{1}{c}{--} & & \multicolumn{1}{c}{--} & \\
\hline\hline

\multicolumn{11}{|c|}{\bfseries SFT (single-task)} \\ \hline

\multirow{5}{*}{\bfseries Unixcoder-base}
& Full     & 100.00 & 94.34 & & 65.19 & & 73.93 & & 33.51 & \\
& Serial adapter   &   1.85 & 94.71 & \UP{0.37} & 63.07 & \DOWN{2.12} & 69.45 & \DOWN{4.48} & 36.68 & \UP{3.17} \\
& Parallel adapter &   0.93 & 94.36 & \UP{0.02} & 64.86 & \DOWN{0.33} & 67.58 & \DOWN{6.35} & 35.31 & \UP{1.80} \\
& LoRA     &   0.46 & 93.84 & \DOWN{0.50} & 62.59 & \DOWN{2.60} & 66.22 & \DOWN{7.71} & 37.28 & \UP{3.77} \\
& Prefix   &   7.16 & 92.75 & \DOWN{1.59} & 62.41 & \DOWN{2.78} & 68.73 & \DOWN{5.20} & 38.10 & \UP{4.59} \\
\hline

\multirow{5}{*}{\bfseries Codet5+ 770M}
& Full     & 100.00 & 94.73 & & 65.37 & & 69.57 & & 35.41 & \\
& Serial adapter   &   1.70 & {93.90} & \DOWN{0.83} & {65.48} & \UP{0.11} & 68.87 & \DOWN{0.70} & {37.13} & \UP{1.72} \\
& Parallel adapter  &   0.85 & {94.41} & \DOWN{0.32} & {64.20} & \DOWN{1.17} & 69.08 & \DOWN{0.49} & {36.47} & \UP{1.06} \\
& LoRA     &   0.42 & {93.99} & \DOWN{0.74} & {63.56} & \DOWN{1.81} & {70.21} & \UP{0.64} & {37.70} & \UP{2.29} \\
& Prefix   &   6.43 & {90.35} & \DOWN{4.38} & 56.30 & \DOWN{9.07} & 56.92 & \DOWN{12.65} & {28.27} & \DOWN{7.14} \\
\hline

\multirow{5}{*}{\bfseries Deepseek coder 1.3B}
& Full     & 100.00 & 93.10 & \ & {62.96} & & 66.41 & & 27.92 & \\
& Serial adapter   &   0.98 & {93.38} & \UP{0.28} & {63.87} & \UP{0.91} & {71.63} & \UP{5.22} & {41.76} & \UP{13.84} \\
& Parallel adapter  &   0.84 & 93.50 & \UP{0.40} & {63.62} & \UP{0.66} & 68.69 & \UP{2.28} & {36.55} & \UP{8.63} \\
& LoRA     &   0.24 & {93.75} & \UP{0.65} & {63.73} & \UP{0.77} & 66.90 & \UP{0.49} & {43.57} & \UP{15.65} \\
& Prefix   & \multicolumn{1}{c|}{--} & \multicolumn{1}{c}{--} & & \multicolumn{1}{c}{--} & & \multicolumn{1}{c}{--} & & \multicolumn{1}{c}{--} & \\
\hline

\multirow{5}{*}{\bfseries Qwen2.5-Coder-1.5B}
& Full     & 100.00 & {94.23} & & {63.18} & & 67.88 & & 29.80 & \\
& Serial adapter   &   0.71 & {94.43} & \UP{0.20} & {62.70} & \DOWN{0.48} & 69.90 & \UP{1.72} & {39.82} & \UP{10.02} \\
& Parallel adapter &   0.35 & \multicolumn{1}{c}{94.08} & \DOWN{-0.15} & \multicolumn{1}{c}{62.77} & \DOWN{0.51} & 70.42 & \UP{2.54} & {37.26} & \UP{7.46} \\
& LoRA     &   0.14 & {93.99} & \DOWN{0.24} & {62.85} & \DOWN{0.33} & 69.82 & \UP{1.94} & {39.81} & \UP{10.01} \\
& Prefix   & \multicolumn{1}{c|}{--} & \multicolumn{1}{c}{--} & & \multicolumn{1}{c}{--} & & \multicolumn{1}{c}{--} & & \multicolumn{1}{c}{--} & \\
\hline
\end{tabular}
}
\label{tab:mft_sft_methods}
\endgroup
\end{table*}

\begin{figure}[!htbp]
  \centering
  \begin{minipage}{0.56\columnwidth}
    \centering
    \includegraphics[width=\linewidth]{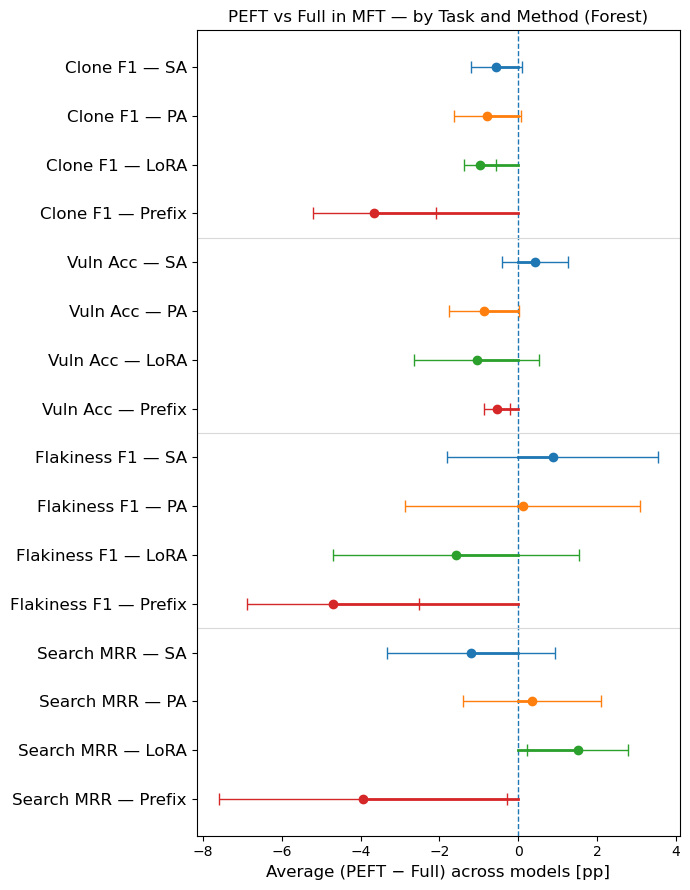}
        \caption{Mean performance difference (PEFT - full fine-tuning) across four models, reported separately for each task–PEFT pair. Dots represent average differences, while horizontal bars indicate 95\% confidence intervals. Colors denote the PEFT method: blue = Serial Adapter (SA), orange = Parallel Adapter (PA), green = LoRA, and red = Prefix. The x-axis shows differences in percentage points (pp), where positive values indicate superior performance of PEFT relative to full fine-tuning. }
    \label{fig:peft_full_mft}
  \end{minipage}\hfill
  \begin{minipage}{0.42\columnwidth}
    \centering
    \includegraphics[width=\linewidth]{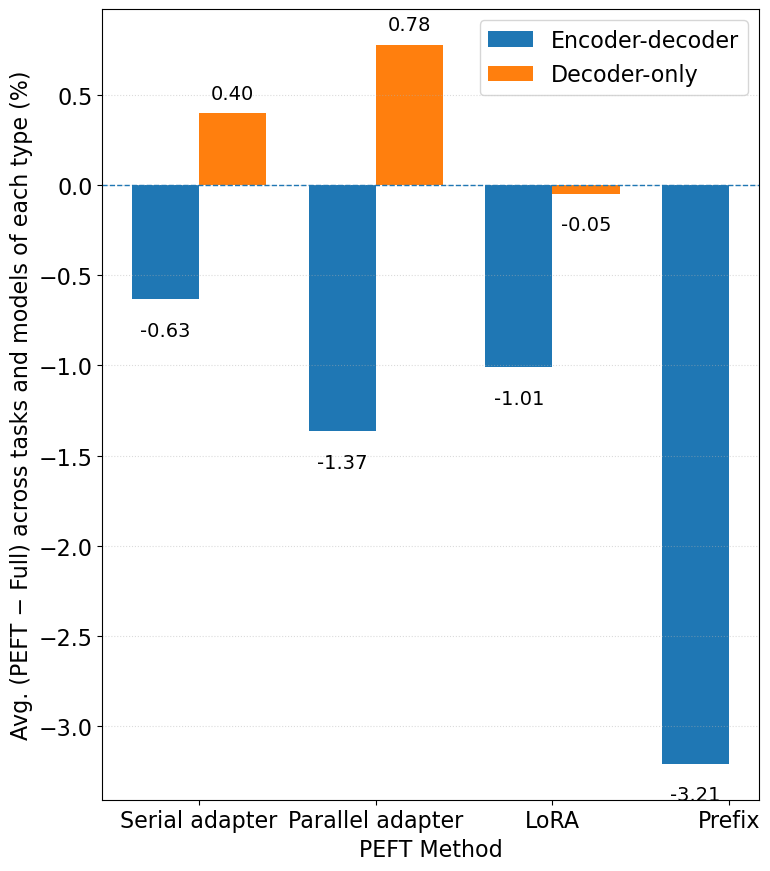}
    \caption{Average performance difference (PEFT - full fine-tuning) across tasks, grouped by model type and PEFT method. Bars show mean differences in percentage points (\%) for encoder-decoder models (blue) and decoder-only models (orange). Positive values indicate performance gains from PEFT relative to full fine-tuning, while negative values indicate losses. }
    
    \label{fig:encoder_decoder}
  \end{minipage}
\end{figure}

To answer RQ1, we compare multiple PEFT approaches against full-model fine-tuning in a multi-task setting to assess whether PEFT methods can retain their effectiveness in multi-task learning, as they have consistently demonstrated in single-task scenarios.

Table \ref{tab:mft_sft_methods} (top part) reports results from fine-tuning each backbone model either jointly on all four tasks, denoted \emph{MFT} (Multi-task Fine-tune), using either full fine-tuning or PEFT. For clarity, we compute the performance difference between each PEFT method and full fine-tuning, and we indicate gains with an upward green arrow and drops with a downward red arrow. Figure \ref{fig:peft_full_mft} also shows the average difference across all models.

\textbf{PEFT for multi-task learning is competitive to full fine-tuning.} For \emph{clone detection} and \emph{vulnerability detection}, PEFT methods are close to be as effective as full fine-tuning: the deltas cluster within about \(\pm 1\) \%, with mostly small negative shifts. The \emph{test flakiness} and \emph{code search} tasks show more variability depending on the PEFT method and type of model used. Flakiness swings by method and model (e.g., gains with adapters on decoder-only, but losses with LoRA/Prefix on encoder-decoder), as detailed below. 

\textbf{Serial and parallel adapters perform best for classification tasks, and LoRA for retrieval.} Adapters deliver small improvements in several cases (e.g., Deepseek code search PA $+2.50$; Qwen test flakiness SA/PA $+4.23/+3.77$).  LoRA is broadly neutral overall but stands out on search, where it is consistently beneficial. 
Prefix-tuning is the weakest across tasks, consistent with the strongly negative bar in the right panel.

\textbf{Decoder-only models benefit most from PEFT: } adapters (and often LoRA) yield small, consistent gains, which explains the positive orange bars. Encoder-decoder models tend to show modest regressions for PEFT, especially Prefix, matching the negative blue bars. Net result: PEFT is effective for multi-task learning—stable and near-zero on clone/vulnerability, while performance on flakiness and search depends on the chosen PEFT method and the underlying architecture.

Overall, in multi-task settings, PEFT methods perform comparably to full fine-tuning and, in some cases, surpass it. LoRA tends to benefit code search because retrieval depends on precise attention representations; low-rank adaptations efficiently reweight the Q/K/V projections and reshape the embedding geometry without destabilizing the network. Clone and vulnerability detection appear stable because the pre-trained backbone already captures the key lexical and structural cues; therefore, PEFT mainly performs minor recalibration of decision boundaries, yielding near-zero deltas. 
Adapters often perform best in multi-task settings by isolating task-specific parameters in bottleneck modules, which reduces gradient interference and negative transfer, an effect especially visible in decoder-only architectures. By contrast, prefix-tuning perturbs only the input context and offers limited capacity to realign internal representations, leading to consistently weaker results.\\

\noindent
\begin{minipage}{\columnwidth}
\setlength{\fboxsep}{6pt} 
\colorbox{gray!15}{
  \parbox{\dimexpr\columnwidth-2\fboxsep-2\fboxrule}{
    \textbf{RQ1 Results.}
PEFT is broadly effective for multi-task learning, often matching, and sometimes outperforming full fine-tuning. 
Serial adapters often perform best for classification tasks, and LoRA is beneficial for retrieval tasks. Gains are strongest on decoder-only models.

  }
}
\end{minipage}


\subsection{RQ2: Multi-task and single-task PEFT}

\begin{figure}[!htbp]
\centering
  \centering
  \includegraphics[width=0.8\linewidth]{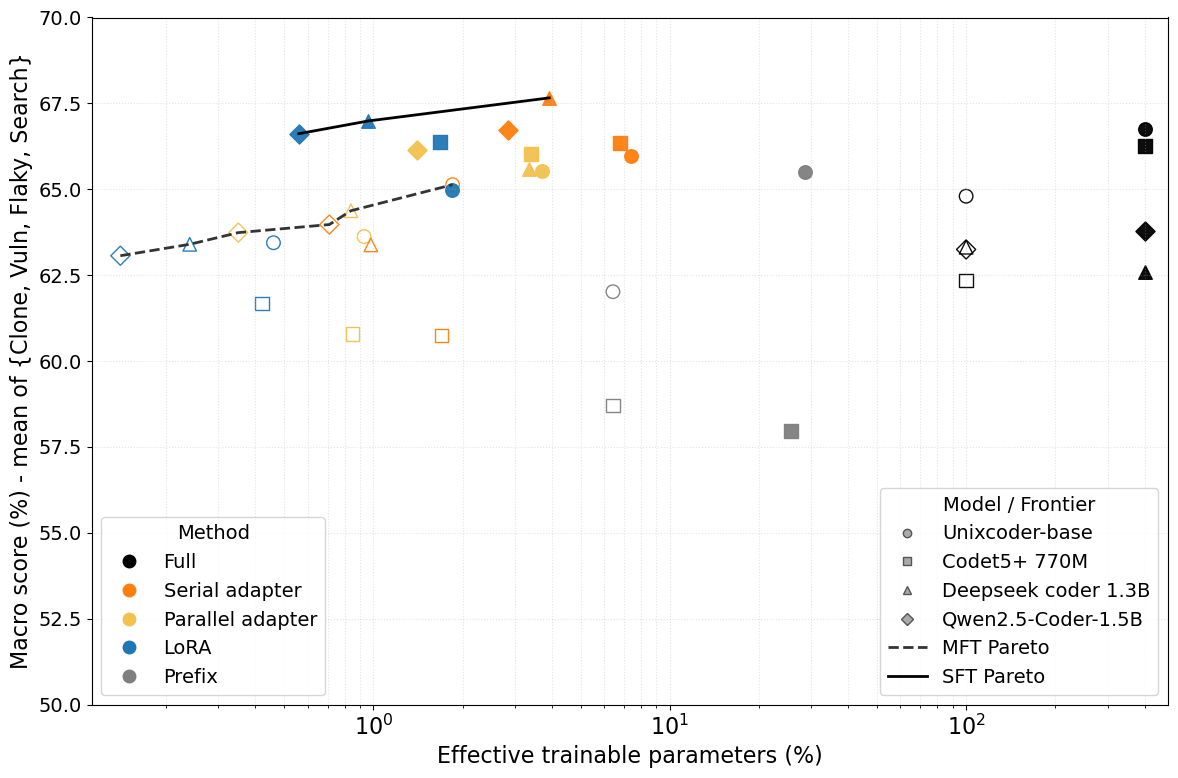}
    \caption{Comparing MFT to SFT: Performance-efficiency trade-off across PEFT methods and models. Each point shows the macro score (mean of \emph{Clone}, \emph{Vuln}, \emph{Flaky}, \emph{Search}) versus the fraction of trainable parameters (relative to full fine-tuning). Colors encode the PEFT method; marker shapes denote the backbone model. Solid and dashed lines trace the empirical Pareto fronts for SFT and MFT, respectively. Higher is better; further left indicates fewer trainable parameters.}
    
  \vspace{-1.0em}
  \label{fig:perf_params}
\end{figure}

\begin{table}[h]
\centering
\small
\caption{ Convergence cost for the Qwen and Deepseek models. We report the tokens processed required to reach the best validation score, computed as \(\text{Updates}\times \text{batch size }\times \text{sequence length }(512)\). The rate column gives the cost ratio \(\text{SFT}/\text{MFT}\); values \(>\!1\) mean SFT is more expensive than MFT. }

\label{tab:qwen_convergence}
\begin{tabular}{l l  r r r}
\toprule
\textbf{Model} & \textbf{Method} &
\multicolumn{1}{c}{\textbf{Tokens}$_{\text{MFT}}$} &
\multicolumn{1}{c}{\textbf{Tokens}$_{\text{SFT}}$} &
\multicolumn{1}{c}{\textbf{SFT/MFT}} \\
\midrule
\multicolumn{5}{l}{\textbf{Qwen}} \\
Qwen & SA   & 257{,}818{,}624 & 1{,}830{,}223{,}872 & 7.10 \\
Qwen & PA    & 644{,}546{,}560 & 1{,}176{,}895{,}488 & 1.83 \\
Qwen & LoRA  & 773{,}455{,}872 & 1{,}686{,}831{,}104 & 2.18 \\
\midrule
\multicolumn{5}{l}{\textbf{Deepseek}} \\
Deepseek & SA   & 1{,}031{,}405{,}568 & 1{,}704{,}656{,}896 & 1.65 \\
Deepseek & PA   & 1{,}031{,}405{,}568 & 1{,}186{,}116{,}403 & 1.15 \\
Deepseek & LoRA &  1{,}031{,}405{,}568 & 3{,}320{,}643{,}584 & 3.22 \\

\bottomrule
\end{tabular}
\end{table}

\begin{figure}[!htbp]
  \centering
  \includegraphics[width=0.8\linewidth]{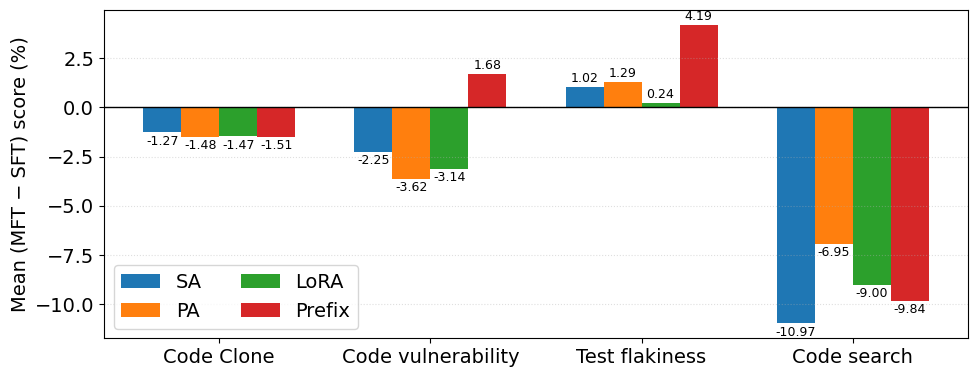}
  \caption{Comparing MFT to SFT: Mean performance difference \((\text{MFT}-\text{SFT})\) by task. Bars show the average score gap across available models (percentage points). Positive values indicate MFT outperforms SFT; negative values indicate the reverse. Colors denote PEFT methods.}
  \vspace{-1.0em}
  \label{fig:sft_mft_bytask}
\end{figure}

This RQ examines the performance–efficiency trade-off of multi-task fine-tuning versus single-task.

\textbf{Multi-task PEFT reduces memory cost.}  
Figure \ref{fig:perf_params} plots macro performance (mean over all tasks) against the fraction of trainable parameters for each model/finetuning method pair. The results show that multi-task PEFT substantially reduces the number of trainable parameters. It remains close to the singl-task PEFT performance frontier while relying on a single PEFT module shared across all tasks. Consequently, for $T$ tasks, the total number of trainable parameters to store is reduced by a factor of $T$ (4$\times$ in our case). In practice, this translates to storing one small adapter/LoRA module instead of four under single-PEFT, or maintaining one model instead of four under full finetuning, leading to simpler deployment when hardware or storage budgets matter more than a minor accuracy drop.


Beyond parameter savings, \textbf{multi-task PEFT also reduces computational cost}. Table \ref{tab:qwen_convergence} summarizes the computation required for Qwen and Deepseek to reach their best validation score under SFT and MFT. We compute the basic computation cost as the number of optimization steps taken until the best validation score is reached, multiplied by the batch size and the sequence length to yield the total number of tokens processed.   The SFT/MFT cost ratio (>1 means SFT costs more) shows that a multi-task PEFT run is substantially cheaper than four single-PEFT runs. For Qwen, the ratios are: Serial Adapter 7.10$\times$ (\textbf{85.9\%} fewer processed tokens), Parallel Adapter 1.83$\times$ (\textbf{45.4\%} fewer), and LoRA 2.18$\times$ (\textbf{54.1\%} fewer). For Deepseek: SA 1.65$\times$ (\textbf{39.4\%} fewer tokens), PA 1.15$\times$ (\textbf{13.0\%} fewer), and LoRA 3.22$\times$ (\textbf{68.9\%} fewer). 
Overall, MFT reduces the computation by approximately 13-69\% on Deepseek and 45-86\% on Qwen relative to SFT, with only minor accuracy/F1 trade-offs.


\textbf{The performance drop in multi-task PEFT is acceptable.}  
Figure \ref{fig:perf_params} shows that the mean performance difference across tasks between SFT and MFT runs with PEFT methods is within $1\%$ to $3\%$. Detailed results per model and PEFT method are available in Table \ref{tab:mft_sft_methods}.  
Figure \ref{fig:sft_mft_bytask} breaks down the score gap $(\text{MFT}-\text{SFT})$ by task. \emph{Code clone} is near-neutral, with small, consistent decreases across methods ($-1.3$ to $-1.5\%$). \emph{Test flakiness} benefits from multitask training, with gains across all PEFT methods ($+0.2$ to $+4.2\%$). \emph{Code vulnerability} shows modest losses for SA/PA/LoRA ($-2.3$ to $-3.6\%$), while Prefix yields a slight gain ($+1.7\%$). \emph{Code search} is the outlier, exhibiting substantial negative transfer across methods ($-7$ to $-11\%$). Thus, the attainable performance/parameter-efficiency trade-off depends on the task mix: clone and flakiness can be jointly optimized with minimal risk, vulnerability exhibits moderate sensitivity, and search is the task most sensitive and negatively affected under joint training. \\

\noindent
\begin{minipage}{\columnwidth}
\setlength{\fboxsep}{6pt} 
\colorbox{gray!15}{
  \parbox{\dimexpr\columnwidth-2\fboxsep-2\fboxrule}{
   \textbf{RQ2 Results.} Multi-task PEFT reduces the number of trainable parameters across $T$ tasks by a factor of $T$, yielding significant storage savings. It also requires less computation to converge compared to single-task PEFT, saving up to $86\%$ of processed tokens for Qwen and up to $69\%$ for Deepseek, thereby lowering the overall training cost. The performance drop is small and task-dependent: near parity for stable tasks, larger for more sensitive ones.

}
}
\end{minipage}


\subsection{RQ3: Pairwise Multi-Task Fine-Tuning Analysis}

\begin{figure*}[!htbp]
\vspace{-1.0em}
  \centering
  \includegraphics[width=1\textwidth,scale=1.0]{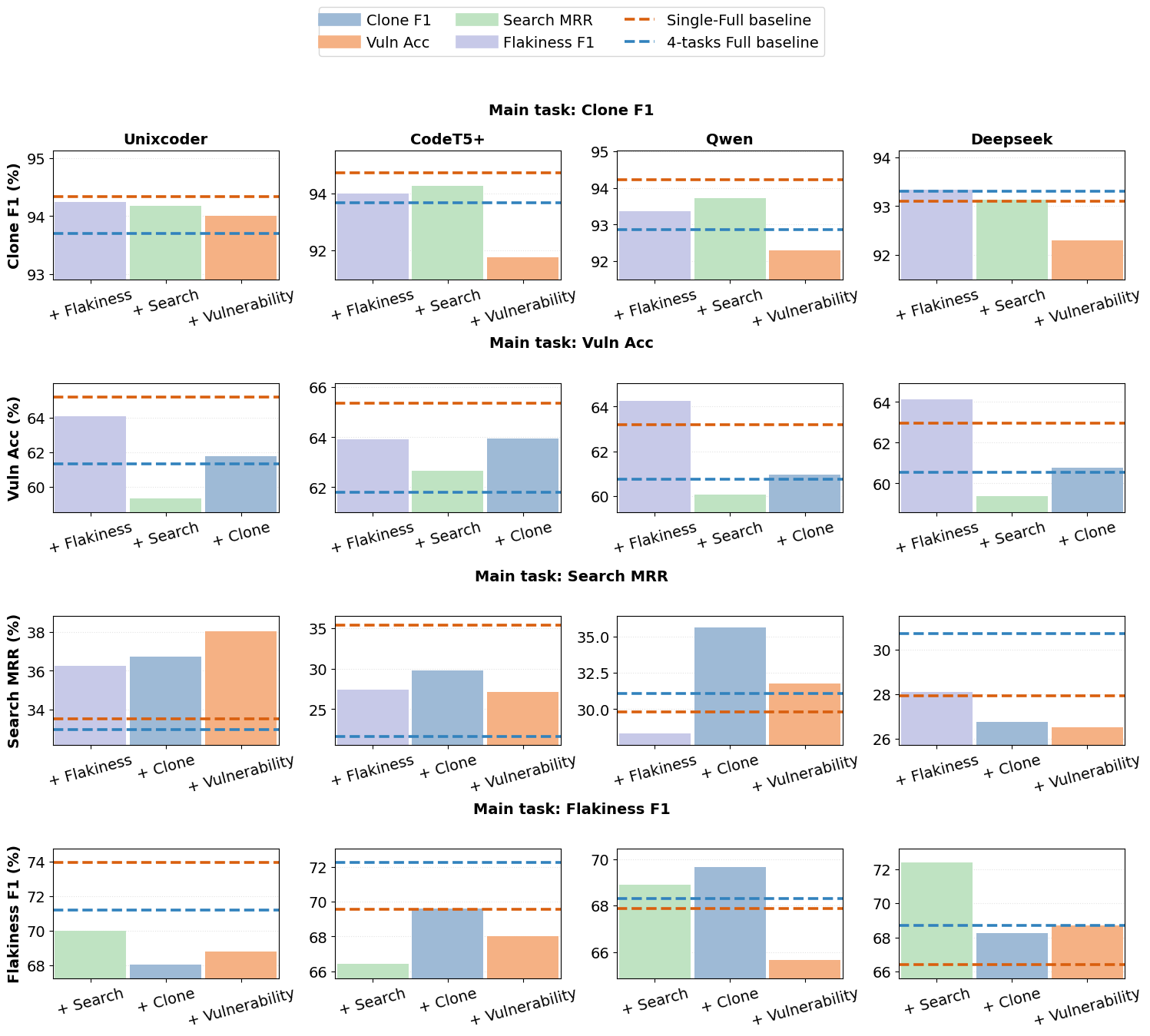}  
\vspace{-0.5em}
    \caption{Pairwise multi-task fine-tuning results. Performance of the four models (Unixcoder, CodeT5+, Qwen coder, Deepseek coder) on multi-task training with the Adapter method. Each row corresponds to a main task (Clone, Vulnerability, code Search, Test Flakiness), while columns show results per model. Bars represent the performance of the main task when it is trained jointly with one of the other three tasks. Dashed lines indicate Single-Full (red) and 4-tasks Full (blue) baselines.}

  \label{fig:pairs}
\end{figure*}

Building on RQ2, which showed that multi-task performance depends on the task mix, we now study pairwise multi-tasking. For each task pair, we jointly fine-tune a single model on both tasks using serial Adapters, which is the most stable PEFT method in RQ1 and 2. Each main task is evaluated when trained together with each of the three other tasks, and Figure \ref{fig:pairs} presents the results of the task pairings in all four models.  The results of pairing are compared against two baselines: separate single-task full models (red dashed lines) and a joint 4-task full model (blue dashed lines).

\textbf{Task stability.}
Some tasks are notably stable across pairings and models. \emph{Clone detection} remains steady when paired with other tasks on all four models, with F1 consistently in the 92--94\% range and usually within $\leq$0.5\% of the single-task baseline. Pairing Clone with Flakiness yields comparable outcomes; pairing with Vulnerability is slightly weaker but tracks the same pattern across models. This stability likely reflects the large, relatively clean dataset and the task’s tight focus on semantic similarity at the code-fragment level. \emph{Vulnerability detection} is also comparatively stable (accuracies typically 62-64\%), except when combined with Search, which tends to pull it down toward the four-task baseline.

\textbf{Task complementarity.}
Some pairings clearly help each other. \emph{Clone detection} and \emph{Code Search} form the strongest complement: both emphasize semantic similarity, and together they surpass the four-task baseline on three models. For example, on CodeT5+ the Search MRR with Clone rises to $\approx$34-35\% versus a four-task baseline near 31\%, while Clone paired with Search stays near its single-task F1 (within $\approx$0.5 \%). A second, weaker but consistent complement appears between \emph{Vulnerability detection} and \emph{Test Flakiness}: on Qwen, pairing with Flakiness lifts Vulnerability to about 63\% (above both the four-task and single-task baselines), with similar gains on Deepseek. In contrast, pairing either Vulnerability or Flakiness detection with \emph{Code Search} usually degrades performance, indicating a mismatch between the representations needed for semantic similarity (Search/Clone) and those for defect signals (Vulnerability/Flakiness).

\textbf{Asymmetry.}
Gains are not always reciprocal. \emph{Vulnerability dection} consistently benefits from \emph{Test Flakiness}, often $+$1-2\% and above the four-task baseline on all models, exceeding the single-task baseline on Qwen and Deepseek, but \emph{Tets Flakiness} does not benefit in return. The likely reason is that the noisier flakiness signal regularizes vulnerability prediction (broadening attention to fault-prone code), whereas the narrower, structured vulnerability signal does little to reduce noise sources in flakiness (environmental instability, concurrency, randomness).

\textbf{Architecture matters.}
Sensitivity to model family is most visible for \emph{Code Search} and \emph{Test Flakiness}. In encoder-decoder models (Unixcoder, CodeT5+), \emph{Search} paired with Clone or Flakiness reaches $\approx$30-35\% MRR and often clears the four-task baseline; in decoder-only models (Qwen, Deepseek), the same pairings are mixed: Qwen improves when paired with Clone, whereas Deepseek underperforms across pairings. \emph{Test Flakiness} shows the opposite pattern: pairings tend to \emph{decrease} F1 by $\approx$2-3\% on encoder–decoder models, but on decoder-only models, pairing with \emph{Code Search} is strongest, reaching $\approx$70-71\% and surpassing both single-task and four-task baselines. By contrast, \emph{Clone detection}  \emph{Vulnerability} maintains similar trends across architectures.

\textbf{Task addition is not always beneficial.}
Pairwise training often outperforms the four-task full setting (blue dashed), but not universally. A clear counter-example is \emph{Flakiness} on Unixcoder, where the four-task configuration reaches $\approx$72\% F1, higher than any pairwise variant (typically 66-70\%). Likewise, \emph{Vulnerability} paired with \emph{Search} frequently fails to match the four-task score across models. These cases suggest that adding extra tasks can disrupt the shared signal of well-matched pairs and that full joint training can sometimes be preferable. \\

\noindent
\begin{minipage}{\columnwidth}
\setlength{\fboxsep}{6pt} 
\colorbox{gray!15}{
  \parbox{\dimexpr\columnwidth-2\fboxsep-2\fboxrule}{
    \textbf{RQ3 Results.} Multi-task performance is shaped by several factors. 
    \textit{Task complementarity:} pairings are most effective when tasks share representational needs, while divergent objectives often degrade performance. 
    \textit{Architecture matters:} a pairing may succeed on decoder-only models but not on encoder-decoder models. 
    \textit{Task stability:} Some tasks, such as Clone detection, remain consistent across pairings, while others, such as Code Search, are more sensitive. 
    \textit{Task addition is not always beneficial: } extra tasks may degrade the main joined tasks.
    \textit{Asymmetry:} benefits are not always reciprocal; a task may gain from another without improving it in return.

  }
}
\end{minipage}

\subsection{RQ4:  direct prompting of larger LLMs }

After investigating the accuracy-efficiency profile of PEFT in multi-task settings, we now ask how these gains compare to a common alternative in practice: zero-shot prompting of large, general-purpose LLMs. In other words, is it better to fine-tune a small code model with a single PEFT module on a group of tasks, or to rely on the capacity of a large LLM? 

Using the prompts listed in the previous section, we evaluate five recent instruction-tuned LLMs in a zero-shot setting on the four tasks, without any task-specific fine-tuning.

\begin{figure}[!htbp]
\vspace{-1.0em}
  \centering
  \includegraphics[width=0.7\columnwidth,scale=1.0]{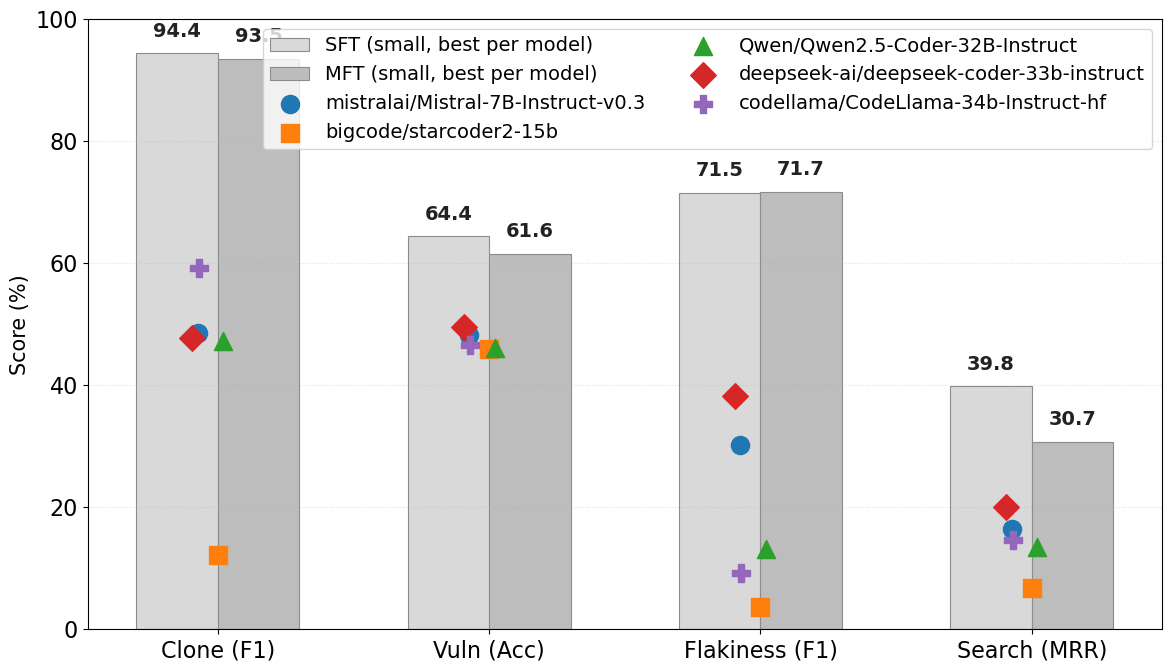}  
\vspace{-0.5em}
 \caption{Zero-shot performance of large instruction-tuned LLMs versus PEFT on compact code models. Bars show the per-task mean of the best fine-tuned scores across small backbones and PEFT methods under SFT and MFT. Markers denote zero-shot performance of large instruction-tuned LLMs. }

  \label{fig:larger_llms}’
\end{figure}

Figure \ref{fig:larger_llms} contrasts these with the average scores of the four small backbones we fine-tuned via PEFT under both SFT and MFT regimes. Both PEFT–SFT and PEFT–MFT decisively outperform zero-shot 7–34B general LLMs across all tasks: Clone reaches 93–94 F1\% (PEFT), while the best zero-shot score is 59.12 F1\%; Vulnerability attains 61–64 Acc\% (PEFT), best zero-shot 49.49 Acc\%; Flakiness reaches 71–72 F1\% (PEFT), best zero-shot 38.25 F1\%; and Search reaches 30–40 MRR\% (PEFT), best zero-shot 20.10 MRR\%.  The results demonstrate that lightweight, task-specific PEFT on compact code models actually outperforms zero-shot prompting of much larger general-purpose LLMs on code analysis tasks.


Beyond accuracy, PEFT-MFT is far more efficient: trains only a small fraction of the parameters and groups all tasks into a single model with a single adapter, yielding large savings in trainable weights, storage, and inference cost compared with repeated zero-shot prompting of 30B\(+\) models.


\noindent
\begin{minipage}{\columnwidth}
\setlength{\fboxsep}{6pt} 
\colorbox{gray!15}{
  \parbox{\dimexpr\columnwidth-2\fboxsep-2\fboxrule}{
    \textbf{RQ4 Results.}
    Multi-task PEFT on compact code models consistently outperforms zero-shot prompting of general-purpose LLMs on code analysis tasks. It achieves performance close to single-task PEFT while substantially reducing trainable parameters, storage, and compute. This makes PEFT–MFT a competitive approach whenever fine-tuning is feasible.
  }
}
\end{minipage}

\section{Threats to validity }
\textbf{External Validity :} Our findings are derived from four code-analysis tasks, covering both classification and retrieval, which may limit their generalizability to other task types. We conduct experiments on well-established CodeXGLUE benchmarks with varying levels of complexity and train four backbone models ranging from 127M to 1.5B parameters. Due to computational constraints (46GB of RAM), the largest model we were able to fine-tune was a 2B-parameter variant. Nevertheless, our selection includes recent and competitive models such as DeepSeek Coder and Qwen Coder, and we also incorporate larger models of up to 34B parameters as baselines for zero-shot inference, thereby diversifying the evaluation. While these choices span diverse architectures and dataset scales, they may not fully generalize to other tasks or to larger or differently pre-trained models. To mitigate this limitation, we focus on widely used tasks and representative backbones. In future work, we plan to extend our evaluation to additional tasks, such as code synthesis and summarization, as well as to a broader range of model families and sizes.

\textbf{Internal Validity:} All models in our experiments were fine-tuned using the same hyperparameter settings and standard values from prior PEFT studies to ensure a fair comparison. Although this approach guarantees consistency, it may not capture the optimal configuration for each model. For instance, the DeepSeek Coder exhibited more severe forgetting, suggesting that model-specific learning rate adjustments could be beneficial. In addition, the datasets vary in size, which could bias training toward larger collections. We mitigate this by employing round-robin sampling (oversampling smaller datasets to equalize task contributions) and applying early stopping based on validation loss. Future work will explore automated hyperparameter optimization and targeted data augmentation strategies to further reduce these risks.

\section{Conclusion }

This paper presents the first systematic evaluation of combining multi-task learning and parameter-efficient fine-tuning for code analysis. We jointly fine-tune diverse code understanding tasks across four code LLMs with distinct architectures, using four PEFT methods: serial adapters, parallel adapters, LoRA, and prefix tuning.

Our study reveals that PEFT is effective for multi-task learning, often reaching parity with full fine-tuning while collapsing several tasks into a single model. Among the methods, serial adapters are the most reliable overall, whereas LoRA is particularly strong for retrieval-oriented objectives, such as code search. Multi-task PEFT also delivers substantial efficiency gains compared to single-task PEFT (dividing the number of trainable parameters by approximately the number of tasks) and to multi-task full fine-tuning (up to 99.8\% training time); and it cuts compute costs by more than 50\%, typically with only small, task-dependent accuracy drops. At the same time, our analysis shows that transfer dynamics are critical; task stability, model architecture, task complementarity, asymmetry, and signal quality jointly shape the success of co-fine-tuning. Finally, we find that multi-task PEFT on compact, code-specialized backbones consistently outperforms zero-shot prompting of much larger general-purpose LLMs on code-analysis tasks, preserving near-SFT accuracy while substantially reducing storage and compute demands.

Taken together, these findings constitute the first comprehensive evaluation of PEFT for multi-task code analysis, distill practical guidelines for method selection and task pairing, and position compact PEFT-MFT models as a cost-efficient alternative to massive general-purpose LLMs. Future work could investigate principled task grouping and stabilization techniques for heterogeneous objectives, including extensions that bridge code understanding and code generation.


\bmhead{Acknowledgements}

This work is supported by the Luxembourg National Research Fund (FNR), Grant number C22/IS/17426831/MeMoRIA. Dr. Papadakis is supported by the Luxembourg National Research Fund (FNR) INTER/MOBILITY/2024/IS/18956086.

\bibliography{references}

\end{document}